\pdfoutput=1
\documentclass[oneside]{article}
\usepackage{crop}

\RequirePackage{datetime}
\RequirePackage[dvips]{graphicx}
\RequirePackage{epsfig}
\RequirePackage{url}
\RequirePackage{xspace}

\RequirePackage{amsfonts}
\RequirePackage{amsmath}
\RequirePackage{latexsym}

\makeatletter
 \renewcommand{\maketitle}{\par
  \begingroup
    \renewcommand{\thefootnote}{\fnsymbol{footnote}}%
    \def\@makefnmark{\hbox to\z@{$\m@th^{\@thefnmark}$\hss}}%
    \long\def\@makefntext##1{\parindent 1em\noindent
            \hbox to1.8em{\hss$\m@th^{\@thefnmark}$}##1}%
    \if@twocolumn
      \ifnum \col@number=\@ne
        \@maketitle
      \else
        \twocolumn[\@maketitle]%
      \fi
    \else
      \newpage
      \global\@topnum\z@
      \@maketitle
    \fi
    \thispagestyle{plain}\@thanks
  \endgroup
  \setcounter{footnote}{0}%
  \let\thanks\relax
  \let\maketitle\relax\let\@maketitle\relax
  \gdef\@thanks{}\gdef\@author{}\gdef\@title{}}

\def\@maketitle{%
  \newpage%
    \vskip -3pc%
    \vskip 48\p@
 \parindent 0pt
\begin{center}
    {\LARGE  \@title \par}%
    \vskip 2pc%
         \lineskip .5em%
       {\normalsize  \@author
            \par}%
    \lineskip .1em%
    {\footnotesize   \@date}%
   \end{center}%
  \par
  \vskip 1pc
}

\makeatother
\newcommand{\sign}{\text{sign}}

\usepackage{textcomp}

\usepackage{listings}

\lstdefinelanguage{GLSL}
{morekeywords={if,else,while,do,return,
float,vec2,vec3,vec4,uniform,varying,
abs,clamp,dot,floor,fract,max,min,mix,mod,smoothstep,step,sqrt},
sensitive=true,
morecomment=[l]{//},
morecomment=[s]{/*}{*/},
morestring=[b]",
}

\definecolor{listinggray}{gray}{1.0}
\definecolor{lbcolor}{rgb}{1.0,1.0,0.9}
\begin{document}
\setcounter{page}{1}
%*******************************************************************************
% Running head
%
% Replace [AUTHOR(S)] with last name of author.  If two authors, separate last
% names by "and" (e.g., Foo and Bar).  For three or more authors, use the first
% author followed by "et al." (e.g., Foo et al.).
%
% Replace [TITLE] with the article title.
%*******************************************************************************
\markboth{journal of graphics tools}{McEwan et al: Efficient computational noise in GLSL}

%*******************************************************************************
% Title / Author
%
% Replace [TITLE] with the article title.
%
% Replace [AUTHOR(S)] with the author.  If two author(s) with the SAME affiliation,
% separate the names using "and" (e.g., John Doe and Jane Doe).  For three or more
% authors with the SAME affiliation, separate names by commas and preceding the
% final name with "and" (e.g., John Doe, Jane Doe, and Foo Bar).
%
% If authors have different affiliations, uncomment the lines for ADDITIONAL
% AUTHORS and AFFILIATION. Format as described above but group authors with
% the same affiliation. If additional affiliations are needed, copy the two
% commented lines for author and affiliation as needed.
%*******************************************************************************
\title{Efficient computational noise in GLSL}

\label{start}

\author{Ian McEwan$^1$, David Sheets$^1$, Stefan Gustavson$^2$ and Mark Richardson$^1$}

\date{$^1$Ashima Research, 600 S. Lake Ave., Suite 104, Pasadena CA 91106, USA\\
$^2$Media and Information Technology,  Link{\"o}ping University, Sweden}

\maketitle

%*******************************************************************************
% ADDITIONAL AUTHORS AND AFFILIATION
%*******************************************************************************
%\vspace*{-4.5pc} % remove extra space after \maketitle

% copy and paste the following two lines as needed
%\noindent {\helR XXX}
%\noindent {\helRE XXX}

%\vspace*{5pc} % manually put back spacing removed from above

%*******************************************************************************
% Abstract
%*******************************************************************************
\begin{abstract}
We present GLSL implementations of Perlin noise and Perlin simplex noise that
run fast enough for practical consideration on current generation GPU hardware.
The key benefits are that the functions are purely computational, i.e. they use
neither textures nor lookup tables, and that they are implemented in GLSL
version 1.20, which means they are compatible with all current GLSL-capable
platforms, including OpenGL ES 2.0 and WebGL 1.0. Their performance is on par
with previously presented GPU implementations of noise, they are very convenient
to use, and they scale well with increasing parallelism in present and
upcoming GPU architectures.
\end{abstract}

%*******************************************************************************
% Article
\pagestyle{myheadings}
%*******************************************************************************

\begin{figure}[ht]
\includegraphics[width=0.99\textwidth]{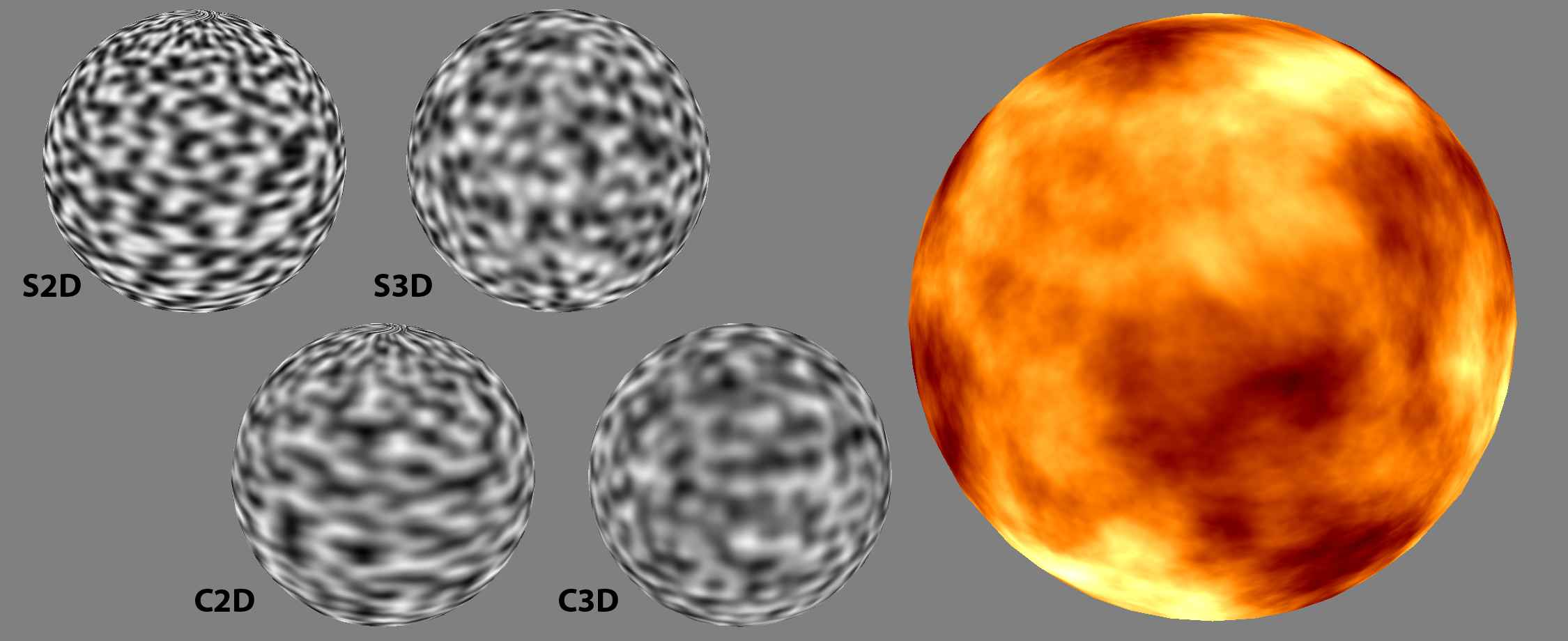}
\caption{\small{2D and 3D simplex noise (S2D, S3D) and 2D and 3D
classic noise (C2D, C3D) on a sphere, and a swirling fire shader
using several noise components.}}
\label{fig:noisedemo}
\end{figure}

\section{Introduction and background}
Perlin noise \cite{Perlin1985,Perlin2002} is one of the most useful building blocks of
procedural shading in software. The natural world is largely built on or from
stochastic processes, and manipulation of noise allows a variety of natural
materials and environments to be procedurally created with high flexibility, at
minimal labor and at very modest computational costs. The introduction of Perlin
Noise revolutionized the offline rendering of artificially-created worlds.

Hardware shading has not yet adopted procedural methods to any significant
extent, because of limited GPU performance and strong real time constraints.
However, with the recent rapid increase in GPU parallelism and performance,
texture memory bandwidth is often a bottleneck, and procedural patterns are
becoming an attractive alternative and a complement to traditional
image-based textures.

Simplex noise \cite{Perlin2001} is a variation on classic Perlin noise, with the
same general look but with a different computational structure. The benefits
include a lower computational cost for high dimensional noise fields, a simple
analytic derivative, and an absence of axis-aligned artifacts.
Simplex noise is a gradient lattice noise just like
classic Perlin noise and uses the same fundamental building blocks. Some
examples of noise on a sphere are shown in Figure~\ref{fig:noisedemo}.

This presentation assumes the reader is familiar with classic Perlin noise and
Perlin simplex noise. A summary of both is presented in \cite{Gustavson2005}.
We will focus on how our approach differs from software implementations and
from the previous GLSL implementations in \cite{Green2005,Gustavson2004}.

\section{Platform constraints}
GLSL 1.20 implementations usually do not allow dynamic access of arrays in
fragment shaders, lack support for 3D textures and integer texture lookups,
have no integer logic operations, and don't optimize conditional code well.
Previous noise implementations rely on many of these features, which
limits their use on these platforms. Integer table lookups implemented
by awkward floating point texture lookups produces unnecessarily
slow and complex code and consumes texture resources. Supporting code outside
of the fragment shader is needed to generate these tables or textures,
preventing a concise, encapsulated, reusable GLSL implementation independent
of the application environment. Our solutions to these problems are:
\begin{itemize}
\item{Replace permutation tables with computed permutation polynomials.}
\item{Use computed points on a cross polytope surface to select gradients.}
\item{Replace conditionals for simplex selection with rank ordering.}
\end{itemize}

These concepts are explained below. The resulting noise functions are
completely self contained, with no references to external data and
requiring only a few registers of temporary storage.

\section{Permutation polynomials}
Previously published noise implementations have used permutation tables or
bit-twiddling hashes to generate pseudo-random gradient indices. Both of these
approaches are unsuitable for our purposes, but there is another way.
A \textit{permutation polynomial} is a function that uniquely permutes a sequence
of integers under modulo arithmetic, in the same sense that a permutation
lookup table is a function that uniquely permutes a sequence of indices.
A more thorough explanation of permutation polynomials can be found in
the online supplementary material to this article. Here, we will only
point out that useful permutations can be constructed using polynomials
of the simple form $(A x^2 + B x) \bmod{M}$. For example, The integers
modulo-9 admit the permutation polynomial $(6 x^2 + x) \bmod{9}$ giving
the permutation $( 0\ 1\ 2\ 3\ 4\ 5\ 6\ 7\ 8 ) \mapsto
( 0\ 7\ 8\ 3\ 1\ 2\ 6\ 4\ 5 )$.

The number of possible polynomial permutations is a small subset of all
possible shufflings, but there are more than enough of them for our purposes.
We need only one that creates a good shuffling of a few hundred numbers, and
the particular one we chose for our implementation is $(34 x^2 + x) \bmod{289}$.

What is more troublesome is the often inadequate integer support in GLSL 1.20
that effectively forces us to use single precision floats to represent integers.
There are only 24 bits of precision to play with (counting the implicit
leading 1), and a floating point multiplication doesn't drop the highest bits on
overflow. Instead it loses precision by dropping the low bits that do not fit
and adjusts the exponent. This would be fatal to a permutation algorithm, where
the least significant bit is essential and must not be truncated in any
operation. If the computation of our chosen polynomial is implemented in the
straightforward manner, truncation occurs when $34 x^2 + x > 2^{24}$, or
$|x| > 702$ in the integer domain. If we instead observe that modulo-$M$
arithmetic is congruent for modulo-$M$ operation on any operand at any time,
we can start by mapping $x$ to $x \bmod{289}$ and then compute the polynomial
$34 x^2 + x$ without any risk for overflow.
By this modification, truncation does not occur for any $x$ that can be
exactly represented as a single precision float, and the noise domain
is instead limited by the remaining fractional part precision for the input
coordinates. Any single precision implementation of Perlin noise,
in hardware or software, shares this limitation.

\section{Gradients on $N$-cross-polytopes}

\begin{figure}[ht]
\includegraphics[width=0.5\textwidth]{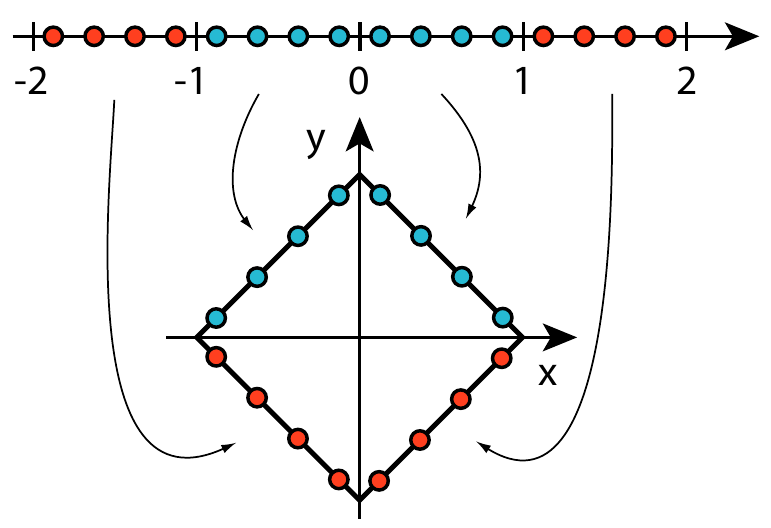}
\caption{\small{Mapping from a 1-D line segment to the boundary of
a 2-D diamond shape.}}
\label{fig:crosspolytope2D}
\end{figure}

\begin{figure}[ht]
\includegraphics[width=0.9\textwidth]{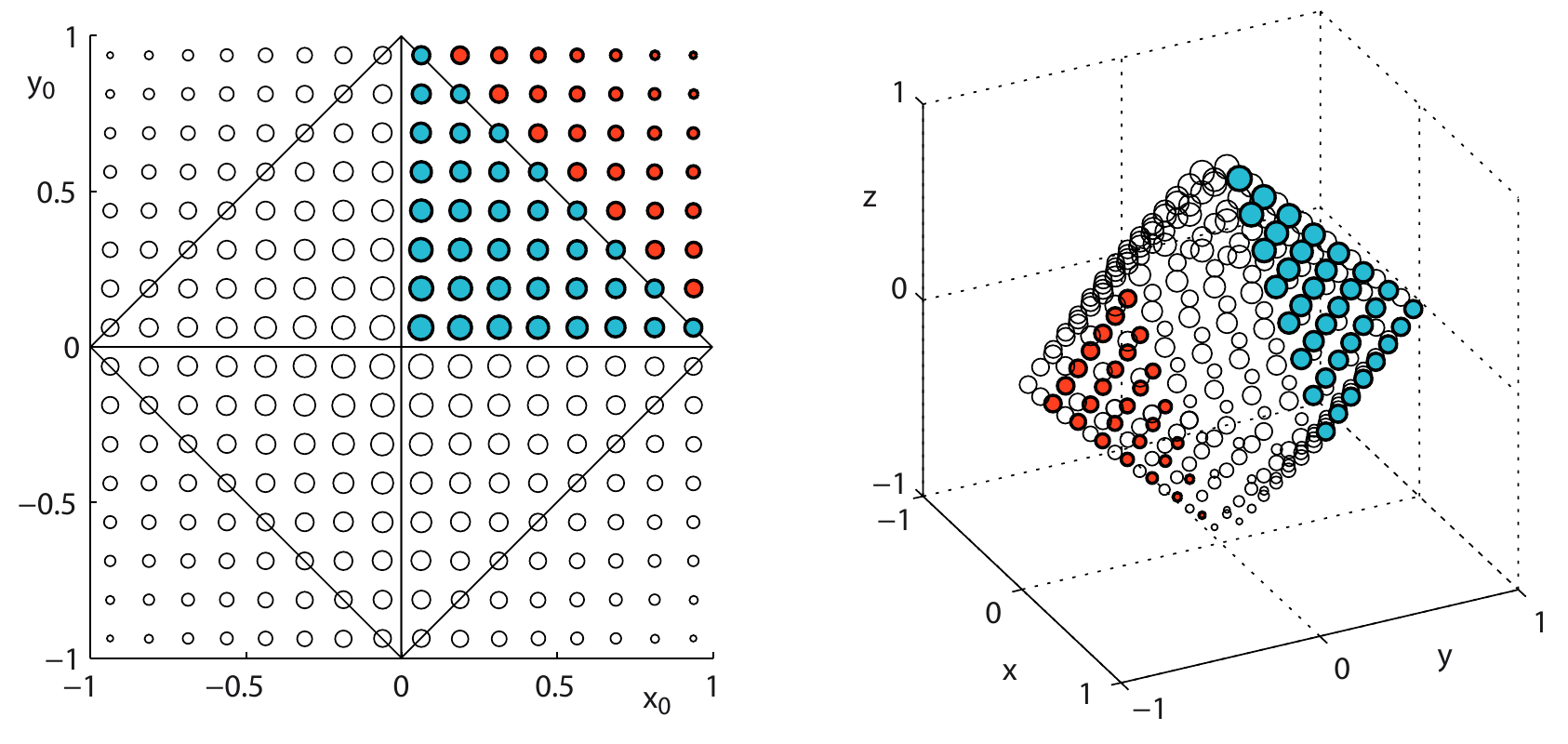}
\caption{\small{Mapping from a 2-D square to the boundary of a 3-D octahedron.
Blue points in the quadrant $x>0, y>0$ where $|x|+|y| < 1$ map to the face
$x, y, z > 0$, while red points where $|x|+|y| > 1$ map to the opposite face
$x, y, z < 0$.}}
\label{fig:crosspolytope3D}
\end{figure}

Lattice gradient noise associates pseudo-random gradients with each lattice
point. Previous implementations have used pre-computed lookup tables or bit
manipulations for this purpose. We use a more floating-point friendly way
and make use of geometric relationships between generalized octahedrons in
different numbers of dimensions to map evenly distributed points from an
($N$-1)-dimensional cube onto the boundary of the $N$-dimensional 
equivalent of an octahedron, an $N$-cross polytope. For $N=2$, points on
a line segment are mapped to the perimeter of a rotated square, see
Figure~\ref{fig:crosspolytope2D}. For $N=3$, points in a square map to an
octahedron, see Figure~\ref{fig:crosspolytope3D}, and for $N=4$, points in
a cube are mapped to the boundary of a 4-D truncated cross polytope.
Equation~(\ref{eqn:mapping}) presents the mappings for the 2-D, 3-D and
4-D cases.

\begin{align*}
\label{eqn:mapping}
\textbf{2-D:\quad} & x_0 \in [ -2, 2 ], \quad y = 1 - | x_0 | \tag{1}\\
&\textbf{if~} y > 0 \textbf{~then~} x = x_0 \textbf{~else~} x = x_0 - \sign(x_0)\\
\textbf{3-D:\quad} & x_0, y_0 \in [ -1, 1 ], \quad z = 1 - | x_0 | - | y_0 | \\
&\textbf{if~} z > 0 \textbf{~then~} x = x_0, ~ y = y_0\\
&\textbf{~else~} x = x_0  - \sign(x_0), ~ y = y_0 - \sign(y_0)\\
\textbf{4-D:\quad} & x_0, y_0, z_0 \in [ -1, 1 ], \quad w = 1.5 - | x_0 | - | y_0 | - | z_0 | \\
&\textbf{if~} w > 0 \textbf{~then~} x = x_0, ~ y = y_0, ~ z = z_0\\
&\textbf{~else~} x = x_0 - \sign(x_0), ~ y = y_0 - \sign(y_0), ~ z = z_0 - \sign(z_0)\\
\end{align*}

The mapping for the 4-D case doesn't cover the full polytope boundary -- it
truncates six of the eight corners slightly. However, the mapping covers
enough of the boundary to yield a visually isotropic noise field,
and it is a simple mapping. The 4-D mapping is difficult both to understand
and to visualize, but it is explained in more detail in the supplementary
material.


Most implementations of Perlin noise use gradient vectors of equal length, but
the longest and shortest vectors on the surface of an $N$-dimensional cross
polytope differ in length by a factor of $\sqrt{N}$. This does not cause any
strong artifacts, because the generated pattern is irregular anyway, but for
higher dimensional noise the pattern becomes less isotropic if the
vectors are not explicitly normalized. Normalization needs only to be 
performed in an approximate fashion, so we use the linear part of a 
Taylor expansion for the inverse square root $1 / \sqrt{r}$ in the 
neighborhood of $r = 0.7$. The built-in GLSL function \texttt{inversesqrt()}
is likely to be just as fast on most platforms. Normalization can even be
skipped entirely for a slight performance gain.

\section{Rank ordering}
Simplex noise uses a two step process to determine which simplex contains a
point $\mathbf{p}$. First, the N-simplex grid is transformed to an axis-aligned
grid of $N$-cubes, each containing $N!$ simplices. The determination of which
cube contains $\mathbf{p}$ only requires computing the integer part of the
transformed coordinates. Then, the coordinates relative to the origin of the
cube are computed by inverse transforming the fractional part of the transformed
coordinates, and a rank ordering is used to determine which simplex contains
$x$. Rank ordering is the first stage of the unusual but classic rank
sorting algorithm, where the values are first ranked and then rearranged
into their sorted order. Rank ordering can be performed efficiently by
pair-wise comparisons of components of $\mathbf{p}$.
Two components can be ranked by a single comparison, three components by three
comparisons and four components can be ranked by six comparisons. In GLSL,
up to four comparisons can be performed in parallel using vector operations.
The ranking can be determined in a reasonably straightforward manner from
the results of these comparisons.
The rank ordering approach was used in a roundabout way in the software
4D noise implementation of \cite{Gustavson2005} and the GLSL implementation
of \cite{Gustavson2004}, later improved and generalized by contributions
from Bill Licea-Kane at AMD (then ATI). The 3D noise of \cite{Gustavson2005}
and Perlin's original software implementation presented in \cite{Perlin2001}
instead use a decision tree of conditionals. For details on the rank ordering
algorithm used for 3-D and 4-D simplex noise, which generalizes to $N$-D,
we refer to the supplementary material.

\section{Performance and source code}
The performance of the presented algorithms is good, as presented in
Table \ref{tbl:benchmarks}. With reasonably recent GPU hardware,
2-D noise runs at a speed of several billion samples per second.
3-D noise attains about half that speed, and 4-D noise is somewhat
slower still, with a clear speed advantage for 3-D and 4-D simplex
noise compared to classic noise. All variants are fast enough to
be considered for practical use on current GPU hardware.

Procedural texturing scales better than traditional texturing with
massive amounts of parallel execution units, because it is not
dependent on texture bandwidth. Looking at recent generations of
GPUs, parallelism seems to increase more rapidly in GPUs than texture
bandwidth. Also, embedded GPU architectures designed for OpenGL
ES 2.x have limited texture resources and may benefit from
procedural noise despite their relatively low performance.

The full GLSL source code for 2D simplex noise is quite compact,
as presented in Table~\ref{tbl:sourcecode}. For the gradient mapping,
this particular implementation wraps the integer range $\{0 \ldots 288\}$ repeatedly
to the range $\{0 \ldots 40\}$ by a modulo-41 operation. 41 has no common prime
factors with 289, which improves the shuffling, and 41 is reasonably
close to an even divisor of 289, which creates a good isotropic
distribution for the gradients.

Counting vector operations as a single operation, this code amounts
to just six \texttt{dot} operations, three \texttt{mod}, two \texttt{floor},
one each of \texttt{step}, \texttt{max}, \texttt{fract} and \texttt{abs},
seventeen multiplications and nineteen additions.
The supplementary material
contains source code for 2-D, 3-D and 4-D simplex noise, classic
Perlin noise and a periodic version of classic noise with an
explicitly specified arbitrary integer period, to match the popular
and useful \texttt{pnoise()} function in RenderMan SL. The source
code is licensed under the MIT license. Attribution is required
where substantial portions of the work is used, but there are no
other limits on commercial use or modifications.
Managed and tracked code and a cross-platform benchmark and demo
for Linux, MacOS X and Windows can be downloaded from the public
git repository \texttt{git@github.com:ashima/webgl-noise.git},
reachable also by:\\
\texttt{http://www.github.com/ashima/webgl-noise}

\begin{table}
  \begin{tabular}{l|r|r|r|r|r|r|r|}
  & Const & \multicolumn{3}{|c|}{Simplex noise} & \multicolumn{3}{|c|}{Classic noise} \\
  GPU & color & 2D & 3D & 4D & 2D & 3D & 4D \\
  \hline
  \emph{Nvidia} \\
%  GF9400M & 864 & 147 & 64 & 28 & 146 & 46 & 16 \\ % MacBook Pro, MacOS X 10.6
  GF7600GS & 3,399 & 162 & 72 & 39 & 180 & 43 & 16 \\ % Desktop system, WinXP
  GTX260 & 8,438 & 1,487 &   784 & 426 & 1,477 &   589 & 255 \\ % Desktop system, Win7
  GTX480 & 8,841 & 3,584 & 1,902 & 1,149 & 3,489 & 1,508 & 681 \\ % Desktop system, Linux
  GTX580 & 13,863 & 4,676 & 2,415 & 1,429 & 4,675 & 2,003 & 906 \\ % Desktop system, Linux
  \hline
  \emph{AMD} \\
  HD3650 & 1,974 & 370 & 193 & 117 & 320 & 147 &  67 \\ % Acer laptop, Win7
  HD4850 & 9,416 & 2,586 & 1,320 &   821 & 2,142 &   992 & 457 \\ % Desktop system, Linux
  HD5870 & 18,061 & 4,980 & 3,062 & 2,006 & 4,688 & 2,211 & 1,092 \\ % Desktop system, Win7
  \hline
  \end{tabular}
  \vspace{3pt}
  \caption{\small{Performance benchmarks for selected GPUs, in Msamples per second}}
  \label{tbl:benchmarks}
\end{table}

\section{Old versus new}
The described noise implementations are fundamentally different from
previous work, in that they use no lookup tables at all. The advantage
is that they scale very well with massive parallelism and are not
dependent on texture memory bandwidth. The lack of lookup tables makes
them suitable for a VLSI hardware implementation in silicon, and they can
be used in vertex shader environments where texture lookup is not guaranteed
to be available, as in the baseline OpenGL ES 2.0 and WebGL 1.0 profiles.

In terms of performance, this purely computational noise is not quite
as fast on current GPUs as the previous implementation
by Gustavson \cite{Gustavson2004}, which made heavy use of 2-D texture
lookups both for permutations and gradient generation. Most real time
graphics of today is very texture intensive, and modern GPU architectures
are designed to have a high texture bandwidth.
However, it should be noted that noise is mostly just one component of a
shader, and a computational noise algorithm can make good use of unutilised
ALU resources in an otherwise texture intensive shader. Furthermore,
we consider the simplicity that comes from independence of external
data to be an advantage in itself.

A side by side comparison of the new implementation against the previous
implementation is presented in Table \ref{tbl:oldvsnew}. The old
implementation is roughly twice as fast as our purely computational version,
although the gap appears to be closing with more recent GPU models
with better computing power. It is worth noting that 4D classic noise
needs 16 pseudo-random gradients, which requires 64 simple quadratic
polynomial evaluations and 16 gradient mappings in our new
implementation, and a total of 48 2-D texture lookups in the previous
implementation. The fact that the old version is faster despite its very
heavy use of texture lookups shows that current GPUs are very clearly
designed for streamlining texture memory accesses.

\section{Supplementary material}
\verb|http://www.itn.liu.se/~stegu/jgt2011/supplement.pdf|

\begin{table}
  \begin{tabular}{l|r|r|r|r|r|r|r|}
  & Const & \multicolumn{3}{|c|}{Simplex noise} & \multicolumn{3}{|c|}{Classic noise} \\
  GPU, version & color & 2D & 3D & 4D & 2D & 3D & 4D \\
  \hline
  \emph{Nvidia} \\
  GTX260 new & 8,438 & 1,487 &   784 & 426 & 1,477 &   589 & 255 \\
  GTX260 old &       & 2,617 & 1,607 & 953 & 3,367 & 1,815 & 921 \\
  GTX580 new & 13,863 & 4,676 & 2,415 & 1,429 & 4,675 & 2,003 & 906 \\
  GTX580 old &        & 7,806 & 4,481 & 2,692 & 8,795 & 3,508 & 1,869 \\
  \hline
  \emph{AMD} \\
  HD3650 new & 1,974 & 370 & 193 & 117 & 320 & 147 &  67 \\
  HD3650 old &       & 665 & 413 & 241 & 871 & 333 & 139 \\
  HD4850 new & 9,416 & 2,586 & 1,320 &   821 & 2,142 &   992 & 457 \\
  HD4850 old &       & 4,615 & 2,874 & 1,524 & 5,654 & 1,926 & 956 \\
  \hline
  \end{tabular}
  \vspace{3pt}
  \caption{\small{Performance of old vs. new implementation, in Msamples per second.}}
  \label{tbl:oldvsnew}
\end{table}

\begin{table}
\begin{lstlisting}
// 2D simplex noise
#version 120
vec3 permute(vec3 x) {
  return mod(((x*34.0)+1.0)*x, 289.0); }
vec3 taylorInvSqrt(vec3 r) {
  return 1.79284291400159 - 0.85373472095314 * r; }
float snoise(vec2 P) {
  const vec2 C = vec2(0.211324865405187134, // (3.0-sqrt(3.0))/6.0;
                      0.366025403784438597); // 0.5*(sqrt(3.0)-1.0);
  // First corner
  vec2 i  = floor(P + dot(P, C.yy) );
  vec2 x0 = P -   i + dot(i, C.xx);
  // Other corners
  vec2 i1;
  i1.x = step( x0.y, x0.x ); // 1.0 if x0.x > x0.y, else 0.0
  i1.y = 1.0 - i1.x;
  // x1 = x0 - i1 + 1.0 * C.xx; x2 = x0 - 1.0 + 2.0 * C.xx;
  vec4 x12 = x0.xyxy + vec4( C.xx, C.xx * 2.0 - 1.0);
  x12.xy -= i1;
  // Permutations
  i = mod(i, 289.0); // Avoid truncation in polynomial evaluation
  vec3 p = permute( permute( i.y + vec3(0.0, i1.y, 1.0 ))
		+ i.x + vec3(0.0, i1.x, 1.0 ));
  // Circularly symmetric blending kernel
  vec3 m = max(0.5 - vec3(dot(x0,x0), dot(x12.xy,x12.xy),
    dot(x12.zw,x12.zw)), 0.0);
  m = m*m;
  m = m*m;
  // Gradients from 41 points on a line, mapped onto a diamond
  vec3 x = fract(p * (1.0 / 41.0)) * 2.0 - 1.0 ;
  vec3 gy = abs(x) - 0.5 ;
  vec3 ox = floor(x + 0.5); // round(x) is a GLSL 1.30 feature
  vec3 gx = x - ox;
  // Normalise gradients implicitly by scaling m
  m *= taylorInvSqrt( gx*gx + gy*gy );
  // Compute final noise value at P
  vec3 g;
  g.x  = gx.x  * x0.x  + gy.x  * x0.y;
  g.yz = gx.yz * x12.xz + gy.yz * x12.yw;
  // Scale output to span range [-1,1]
  // (scaling factor determined by experiments)
  return 130.0 * dot(m, g);
}
\end{lstlisting}
\caption{\small{Complete, self-contained source code for 2D simplex
noise. Code for 2D, 3D and 4D versions of classic and simplex noise
is in the supplementary material and in the online repository.}}
\label{tbl:sourcecode}
\end{table}

%\vspace{24pt} \noindent\textbf{Acknowledgments.}\small \quad [ACKNOWLEDGEMENTS]

%*******************************************************************************
% References
%
% Replace [BIBLIOGRAPHY] with \bibitem entries.  If you are familiar with the
% use of BibTeX, you can optionally use the akpbib.bst file provided.
%*******************************************************************************

%*******************************************************************************
% Web Information
%
% Copy and paste the lines for author contact information for each author of
% the article and replace the place holders as indicated.
%*******************************************************************************
%\section*{Web Information:}\small
%[WEB INFO]

% author contact information
\vspace*{12pt}\noindent Ian McEwan, David Sheets and Mark Richardson,
Ashima Research, 600 S. Lake Ave., Suite 104, Pasadena CA 91106, USA
\newline (ijm@ashimaresearch.com, sheets@ashimaresearch.com, mir@ashimaresearch.com)

\vspace*{12pt}\noindent Stefan Gustavson, Media and Information Technology,
ITN, Link{\"o}ping University, 60174 Norrk{\"o}ping, Sweden
\newline (stefan.gustavson@liu.se)

% [DATE]s will be filled-in by managing editor during final production.
% Please ignore.
\vspace*{24pt}\noindent Received [DATE]; accepted [DATE].

\label{end} % DO NOT REMOVE. Provides \pageref for determining last page number.
 % M pages

\end{document}